\documentclass[default,iicol]{sn-jnl}

\jyear{2021}%

\theoremstyle{thmstyleone}%
\theoremstyle{thmstyletwo}%

\theoremstyle{thmstylethree}%
\usepackage{blindtext}
\usepackage{algorithm}
\usepackage{amsmath}
\usepackage{algpseudocode}
\usepackage{subcaption}

\begin{document}

\title[Chic]{CHIC: Corporate Document for Visual question Answering }


\author*[1,2]{\fnm{Ibrahim} \sur{SOULEIMAN MAHAMOUD}}\email{ibrahim.souleiman\_mahamoud@univ-lr.fr}

\author[2,3]{\fnm{Mickaël} \sur{Coustaty}}\email{mickael.coustaty@univ-lr.fr}

\author[1,2]{\fnm{Aurélie} \sur{Joseph}}\email{aurelie.joseph@getyooz.com}

\author[1,2]{\fnm{Vincent} \sur{Poulain d'Andecy}}\email{vincent.poulaindandecy@getyooz.com}

\author[1,2]{\fnm{Jean-Marc} \sur{Ogier}}\email{jean-marc.ogier@univ-lr.fr}

\affil*[1]{\orgdiv{L3i}, \orgname{La Rochelle Université}, \orgaddress{\street{Avenue Michel Crépeau}, \city{La Rochelle}, \postcode{17000}, \country{France}}}

\affil[2]{\orgdiv{Research Teams}, \orgname{Yooz}, \orgaddress{\street{1 Rue Fleming}, \city{La Rochelle}, \postcode{17000} \country{France}}}


\abstract{

The massive use of digital documents due to the substantial trend of paperless initiatives confronted some companies to find ways to process thousands of documents per day automatically. To achieve this, they use automatic information retrieval (IR) allowing them to extract useful information from large datasets quickly.
In order to have effective IR methods, it is first necessary to have an adequate dataset. Although companies have enough data to take into account their needs, there is also a need a public database to compare contributions between state-of-the-art methods. 
Public data on the document exists as DocVQA\cite{DocVQA} and XFUND \cite{xu-etal-2022-xfund}, but these do not fully satisfy the needs of companies. XFUND contains only form documents while the company uses several types of documents (i.e structured documents like forms but also semi-structured as invoices, and unstructured as emails). Compared to XFUND, DocVQA has several types of documents but only 4.5\% of them are corporate documents (i.e invoice, purchase order, etc). All of these 4.5\% of documents do not meet the diversity of documents required by the company.
We propose CHIC a visual question-answering public dataset. This dataset contains different types of corporate documents and the information extracted from these documents meet the right expectations of companies. 

}




\maketitle



\section{Introduction}

\begin{figure*}[ht]\centering 
	\includegraphics[width=\textwidth]{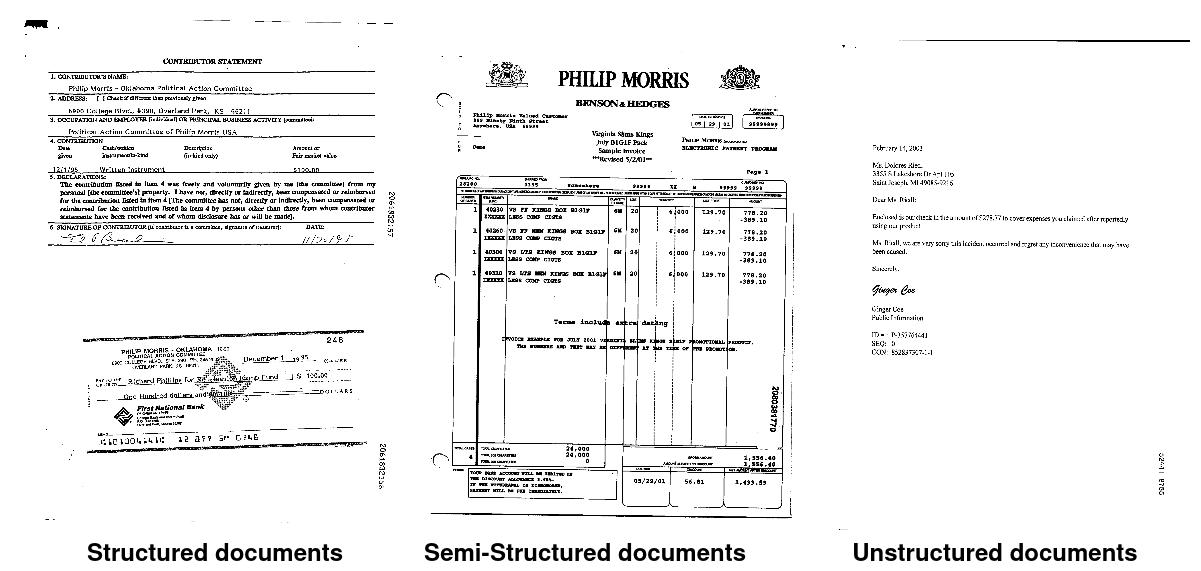}
	\caption{The different types of documents we deal with in our context}
	\label{fig:type_document}
\end{figure*}

Private companies are faced with a huge amount of documents to process every day. These documents come from various sources (e.g internal company processes, subcontractors' documents, and administration documents). Processing such a large amount of documents requires a lot of effort and manpower if an automatic system is not used.
This is where some companies such as YOOZ, Esker, and others are involved to provide automatic document processing solutions. These solutions use Information retrieval (IR) to allow the classification and extraction of useful information from documents in large databases. This permits companies to save both time and money. In order to provide a successful IR solution, it must first be verified that it meets the constraints of the industry.  \newline

These industrial constraints correspond to real needs and bringing solutions to these constraints will allow efficient solutions for the final users.
Among these constraints, we can cite, for example, constraints linked to the speed of processing a document, which must be less than a few seconds. This allows the end-users to have fast results because if the automatic processing of a document takes a few minutes, doing it manually would take the same time and it would not be efficient.  
These IR solutions have to be fast while keeping real efficiency because the quality of the results should not be lost at the expense of saving process time. It is also important to get the most correct results as possible because manually correcting these errors would require more effort and time.  IR solutions should provide results when a high level of confidence is reached, else inform the end-user to process this document. Manual annotation of some cases is faster than checking all the extracted information, where the error is, and correcting it. Once all these constraints are met, it is also necessary to have adequate public data to compare the performance of state-of-the-art models with each other and to have metrics on whether these constraints are respected. \newline

\label{P_industrial_constraints} Public data is very useful for IR methods, as it allows to compare results between different methods, and also to see the limitations of these methods on this type of data. Two aspects characterized this data, the first being the type of documents and the second the elements to be extracted. 
The companies that provide IR solutions are specialized in one type of document because processing any type of document and extracting any type of information would require several platforms, several methods, and considerable effort.  In this paper, we will focus on corporate documents such as (invoices, purchase orders, etc), we have illustrated these different documents in the figure\ref{fig:type_document}.
We can categorize them into three categories. Structured documents (e.g ID, check, form)  are characterized by a high degree of inter-similarity in the layout of the documents, as documents of the same type will always have the same layout. The semi-structured documents (such as invoices, and purchase orders) can change the structure from one document to another but respect a particular organization of the elements that compose them. Finally, the unstructured document (like e-mails) do not respect a predefined layout, and that varied greatly between them.  In the following, we will refer to the set of documents presented above as corporate documents. 
In addition to different types of documents, these companies are confronted with documents in several languages.  This multi-language aspect is required because the companies providing IR solutions work with several companies from different countries and therefore of different languages.
We will focus on European languages, such as English, German, Italian, French, and Spanish. The choice of languages corresponds to the vast majority of documents processed by these companies.
The relevant information to be extracted from these documents for the company is various. In order to take into account all types of information to be extracted, we have privileged the Visual Question Answerings (VQA) approach. These approaches have the advantage of extracting a lot of information with questions and are easy to use by end users.\newline

In the state-of-the-art, we find datasets such as DocVQA and XFUND.
We will explain in the section related-work \ref{related-work}, their limitations (DocVQA, XFUND), and how CHIC is a contribution compared to this state-of-the-art dataset. 

CHIC is a new visual question-answering dataset able to meet the needs of these companies and provide them with a dataset that contains documents they use every day, and the extracted information is similar to the expected one.

Our contributions are summarized as follows:
\begin{itemize}
        \item We propose a new dataset containing 774 corporate documents with 3384 annotated question-answering.
        \item CHIC is the first corporate dataset containing multilingual documents. We have documents in English, German, French, Italian, and Spanish.
        \item CHIC is both a powerful dataset for these documents and also for the questions and answers within. This dataset takes into account the needs of these companies to have diversified questions to extract any type of relevant information.

\end{itemize}

\section{Related-work}\label{related-work}

In this section, we will describe the existing state-of-the-art data. We will first explain each of them and then explain their limitations in our context. 

\subsection{DocVQA}
DocVQA is the first dataset available for questions and answers on company documents. 
DocVQA is a dataset containing 12 767 documents of different types and layouts. These documents include, for example, handwritten documents, forms, advertisements, etc. 
To extract information from the document, they annotated 50,000 questions and answers.
This dataset is perfectly adapted to extract any type of information from any type of document.
Despite the significant contribution of DocVQA with the number of questions and answers available and the documents, there are nevertheless limits to its use in an industrial context.\newline

The limitation of Docvqa can be summarised in two points. This first point is that it has very few corporate documents (e.g. invoices, purchase orders, etc.) present in DocVQA. To arrive at this conclusion, we searched for the type of documents that would correspond to our context. We extracted the corporate documents in DocVQA manually, in order to have higher precision and  this required us 20 man-hours to finalize this task. We found 586 documents that can be used in the corporate context, we will call this subset DocVQA-Corporate. Although DocVQA-Corporate represents 4.5\% of all the documents contained in Docvqa, this is very little.  On the other hand, this number is sufficient to constitute a question and answer dataset on corporate documents. However, after analysing the types of documents present in DocVQA-Corporate, we found that they did not satisfy all the requirements necessary to provide a public dataset in the context of corporate documents. The DocVQA-Corporate contains documents with almost the same layout type, so not enough inter-document diversity. We observed that he had approximately 20 different document layouts, which were duplicated several times in DocVQA-Corporate.This observation led us to conclude that DocVQA-Corporate documents are far from being an appropriate dataset for an industrial context, as it does not take into account the need for great diversity that might occur in an incoming flow of corporate documents. 
The second point concerns the information extracted in DocVQA-Corporate. When we examine the type of responses retrieved from Docvqa-corporate, we find that this search information is primarily (~40\% of responses) strings. Although strings are included in the information sought by companies, the percentage of this type of response is still less than 20\% (i.e when reviewing client information extraction expectations).  These string responses in Docvqa-Corporate are of little or no interest to companies (i.e. only 38\% of these string responses are useful in our cases).  
The rest of the questions correspond well to the expectations of the companies (i.e. we find information on amounts, reference numbers, etc.).
In addition to these two main points, other limitations exist. The first is that DocVQA documents are only in English, whereas in the industrial context often encounter multi-language documents.  Secondly, the annotated answers in DocVQA contain only the text and not the precise location of the answer. Being able to locate an answer very precisely allows the algorithms to learn the correct position of an answer and not simply return the first occurrence found.  This is necessary because in many cases the answer may be found in several places in the document and the algorithms usually select the first answer found as the correct answer and learn from that (e.g. we may have 10 as the total amount, or 10 on a date or 10 may correspond to another number). 
These limitations described above show us that neither DocVQA nor DocVQA-Corporate perfectly meets all the constraints for meeting the need for a public dataset in the industrial context.

\subsection{XFUND}

XFUND is a dataset of contains 1393 document forms (structured documents) containing key-value pairs annotated by humans. In this dataset, key-value means for example, if we have in the document "date of birth: 13 January 2023", in this case, the word date of birth will be annotated as a key and  linked to the value 13 January 2023. The number of annotated key-value is about 30,000 out of the 1393 documents.  These value keys are extracted in several documents in 7 languages (Chinese, Japanese, Spanish, French, Italian, German, and Portuguese).  Despite the real contribution of XFUND, there are several limitations that prevent us from using it in a corporate documents context \newline

We can summarise the limitations of XFUND firstly on its document type but also on the key-value information annotated in this document. 
The first point concerns the XFUND documents, these documents are all structural documents. These structured documents as their name indicate contain words often organized in the same way (i.e. in value keys). Although the company deals with form documents, it also deals with other types of documents (i.e. as seen in the introduction) semi-structured documents, and non-structured documents. In order to respect the types of documents encountered in their daily flow, it is relevant to have documents with several types of structures. Unfortunately, XFUND does not respect the diversity constraint expected by the end users of a public dataset. 
The second point concerns the key-value information extracted in XFUND.  In the annotations done in XFUND, either a word is a key, a value or a title and everything else is categorized as Other. Among these words annotated as Other, we find many words that could be of interest to companies (e.g. a form number that is found without a key at the top). Unfortunately, if companies wanted to use XFUND to extract some Other words, they would have to invest time in annotating them. 
All these limitations together make  the XFUND not a corporate dataset but a dataset of forms for specific needs.

\subsection{RVL-CDIP}

RVL-CDIP is a dataset containing 400,000 documents separated into 15 classes (e.g letters, advertisements, file folders, forms, emails, handwritten, and invoices). Among the annotated classes we find classes that reflect the type of documents often processed in the company but also other types of documents that exist in RVL-CDIP. Although this dataset has corporate documents, there is one limitation.\newline 

In the RVL-CDIP  is intended to be used for document classification. Although document classification is a challenge for companies, the state-of-the-art has a large amount of public data to address this challenge (\cite{RVL-CDIP},\cite{DocClassification}).
In IR, it is certainly relevant to have in the first documents that reflect the context that interests us but it is also important to extract useful information from these documents.
Unfortunately, this limitation means that the documents cannot be used in their current state and would require annotation to be usable. 

\section{CHIC}

\subsection{Formulation of questions and answers problem }
In the state-of-the-art we find several problems related to the question-answers, we can quote question-answers on natural scene image \cite{VQA-Image}, question-answers on text \cite{VQA-Text} and there are question-answers on videos \cite{VQA-Video}. We will describe in this section the problem formulation for predicting the answer to a question in the context of corporate documents.  
We will focus on giving an answer \textbf{a} from a question \textbf{q} and extracting this information from the document \textbf{d} according to a model parameter a method (see the equation \ref{eq:question-answer} ). 
In the following, we describe how we represent $q, d, a$. \newline

The question \textbf{q} is a sequence of words, i.e. $q=[q_1,…,q_n]$, where each $q_t$ is the \textit{t}-th word question with $q_n=$'?'  mark the end of the question. The questions are related in context, it is the documents in our case. In our scenario, we use the image $i$ and the text $t$ of the document $d$. If the original document is a pdf, we convert it to an image, and from this image we extract the text with Optical Character Recognition (OCR) tools. The answers are text that is extracted from the text of the document, an answer can be composed of one word or more. In addition to the text, a response contains the exact position in the image. Thus an answer is composed of several words and a bounding box $a = ([a_1,...,a_2], bounding\_box)$. 

\begin{equation}\label{eq:question-answer}
     \hat{a}= \text{arg max}_{{a}\in \mathcal {A}}  P(a \lvert x,q;\theta) 
\end{equation}

After defining the questions, the documents, and the answers. We have seen a description of CHIC in the following.

\subsection{Overview of the new CHIC dataset }

The CHIC dataset contains 774 multilingual documents in 5 languages (English, French, German, Spanish, and Italian). To be more precise, we have 774 images and text extracted from these images. We extracted the text with the help of Abby FineReader OCR.  The image in CHIC have a high resolution of 2560 * 3332 pixels. 
CHIC imitates as well as possible the different types of documents that could be found in companies, it contains structured, semi-structured, and non-structured documents. 
Then from these documents, we have annotated 3384 questions and answers. In the following, we will explain the steps that allowed us to annotate CHIC.

\subsection{Protocol for annotation}
\label{protcole_annotation}
To be able to build a document that meets the requirements of the company, we first have to find documents that can correspond to this context.
Unfortunately in the state-of-the-art either we have XFUND datasets containing one  category of the document  but a multilingual document. We have also RVL-CDIP which contains a wide variety of documents. Finding the right documents requires taking advantage of the best of both types to build our new dataset.
The documents we use are extracted from two public databases RVL-CDIP\cite{RVL-CDIP} and XFUND\cite{xu-etal-2022-xfund}. The advantage of using them will be twofold. The first is that RVL-CDIP will provide us with semi-structured or non-structured documents all with a different layout from each other, unlike DocVQA-Corporate. 
The second point is that the documents extracted from XFUND are multi-language forms, unlike RVL-CDIP which contains only English documents. Overall we collected 770 documents containing CHIC 655 from RVL-CDIP and 115 from XFUND. In the following, we will be interested in the information extracted from these documents. \newline

The protocol for annotating the questions and answers was as follows. First, we annotated the questions, then the answers to these questions, and finally we checked to see whether the responses are as expected.  In order to annotate the questions, it is first necessary to know what type of questions the end-users need. For this, we have collected the expectations of the type of information to be extracted and what type of question to be used. The instructions were to diversify the questions, in addition to the standard questions  (e.g. what is the net amount, when the word net amount is present in the documents) to add other types of questions.  We formulate some questions to require understanding beyond the content of the document, such as "what is the amount at the top of table I)" or "what is the number at the top of the document reference". We also have questions that require arithmetic and comparison operations on elements of the document, such as "what is the name of the element with the maximum amount in the table?". These other types of questions target information that could not be extracted with standard questions.
After annotating these questions, we then annotated the answers to these questions. The annotation of these questions was done in a special way, we asked each annotator to annotate questions from other annotators and not his own. This will allow confront the understanding of another person on your questions and from this understanding to extract the necessary information. Once the questions and answers were completed, we asked each annotator  to check whether the answer to their question matched their expectations. If the answer did not meet their expectations, we either tried to reformulate the question or find the right answer together. All these annotations could be done by a tool that we have developed ourselves.\newline

In order to perform all these annotations, we did not find any annotation tools that meet our needs, so we built our own annotation tool. The aim of this tool is to be easy to use and intuitive.  For us, we have included only the essential parts. One part allows us to see the image of the document, another part allows us to annotate the questions and their answers. The annotation of the answer includes the answer in the text and also its position in the document in the bounding box.  This annotation tool is available in \cite{chic}.

\begin{figure}[ht]\centering 
	\includegraphics[width=0.5\textwidth]{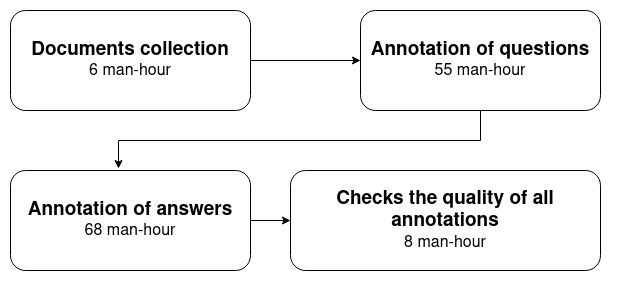}
	\caption{The steps of CHIC annotation and the time it needed }
	\label{fig:steps_document}
\end{figure}

We have summarised in figure \ref{fig:steps_document} the essential states and how much manpower this will require.  We spent a total of approximately 137 man-hours to provide CHIC.

\subsection{Statistics and Analysis}
In this section, we will analyze whether CHIC meets the industrial constraints cited in the \nameref{P_industrial_constraints}. 

We will first start with some analysis of the CHIC documents. We have seen in the sub-section that the documents we have collected come from different databases and that their structures correspond to the expectations of the companies. Corporate documents are different from other types of documents both in their layout and in the words they contain. We  have observed that  93\% of our documents CHIC contain between 100-400 words and only approximately 30 documents contain more than 500 words. When we compare these statistics with another type of document, for example, scientific documents \cite{distrubution_words_paper}, which contain on average 1500 words per document. This shows that the corporate documents contained in CHIC, the words that compose it despite containing on average 150 words. These are distributed over the whole document and not just organized in word blocks (like newspaper articles, scientific papers, etc). In the following, we will go into detail about these words to see which ones are extracted and what their usefulness is.
 \newline

\begin{figure}
  \centering
  \begin{tabular}{@{}c@{}}
    \includegraphics[width=.8\linewidth,height=100pt]{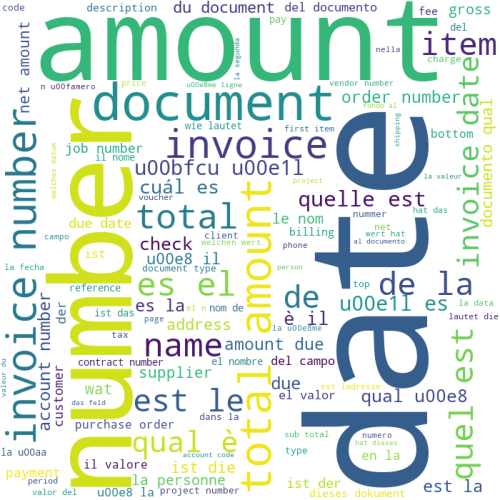} \\[\abovecaptionskip]
    \small (a) Word clouds of  question
  \end{tabular}

  \vspace{\floatsep}

  \begin{tabular}{@{}c@{}}
    \includegraphics[width=.8\linewidth,height=100pt]{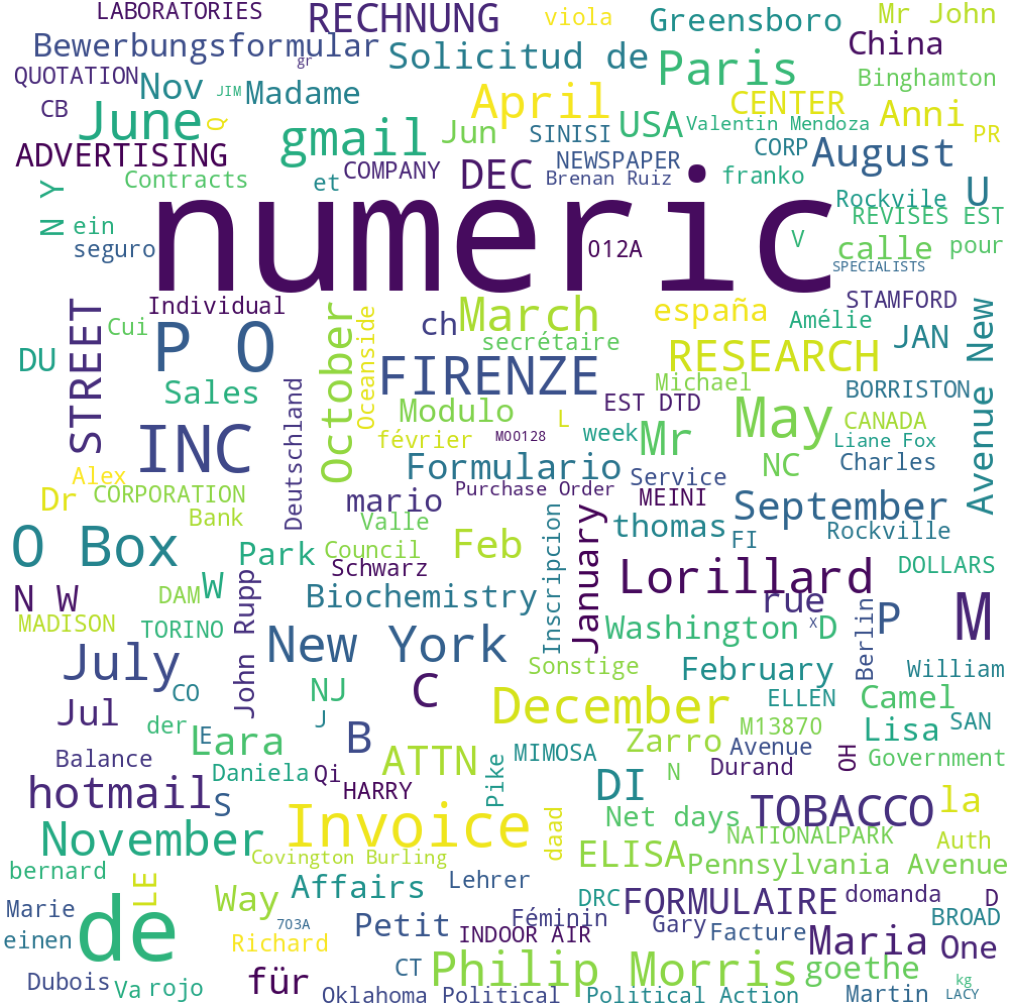} \\[\abovecaptionskip]
    \small (b) Word clouds of  answer
  \end{tabular}

  \caption{}\label{fig:questionanswer_distrubution}
\end{figure}




Figure \ref{fig:word_question_answer}, shows the distribution of the words, both those extracted from the document and the words in the questions. Firstly, we start with the questions words.
The questions in CHIC are mostly related to the date or amount and any other number that concerns the primary needs of the end-users. 
These elements are essential in a company because they are the key information in a document. Being able to extract this information brings many benefits. For example to automatically find the invoice with a precise date or precise amount. Also, they give the information related to the amount to the accountant in order to establish their accounting easily. Many other uses are possible. From these questions, we will extract the answers in the document. 
  
For analysis of the distributions of the answers, the cloud of words does not take into account the numeric, and alphanumerical answers because they are very variable between them. To represent this type of answer we have transformed all the answers of this type into "numeric" words. When we look at the scatter plot we can see that most of the answers are either numeric or alphanumeric, we also have among the answers the type of documents, the name of the company, certainly the supplier of these documents, etc. 

In summary, in CHIC we have various questions of different types and the answers to these questions meet the expectations of the end-users.


\section{Experiments}
In this section, we will explain the methods chosen to provide a CHIC baseline then we explain the metrics used and finally analyze the results obtained.

\subsection{ VQA Models}

To evaluate the performance of the state-of-the-art methods on the CHIC dataset, we have taken two types of methods  pre-trained and non-pre-trained as a baseline. 
 The majority of methods can be divided into two categories: methods that pre-train on the data in an unsupervised way in order to acquire as much prior knowledge as possible (i.e. inspired by human learning mechanisms, which, in order to learn new knowledge, always rely on acquired prior knowledge). Other methods do not use this prior knowledge but are very good at exploiting as much of the current data as possible and extracting as much knowledge as possible in order to achieve a good performance. 
For the pre-trained methods, we have selected Donut \cite{Donut}, LayoutLMv3 \cite{LayoutLMv3}, and LayoutLMv2 \cite{LayoutLMv2}. For the non-pre-trained methods, we selected QALayout \cite{qalayout} and QANet \cite{qanet}.
These two types of methods will allow us to see how important it is to have pre-trained or not-pre-trained methods on CHIC.\newline

LayoutLM-based are pre-trained  methods like LayoutLMv2\cite{LayoutLMv2} or LayoutLMv3\cite{LayoutLMv3}. They are proposed by Microsoft to be used in several tasks of document understanding (document classification, question-answer, etc). Their pre-training is based on the textual, image, and layout features (i.e. the positions of the words in the document). 
The goal of the pre-training is to hide a specific part of the document and find it within the context. The difference between LayoutLMv2, and LayoutLMv3 is essential in the method of masking and the input features used. 
We will first start to explain LayoutLMv2.  The inputs to the model are the text extracted from the image, the image features and layout information related to the word's position in the document. The image features used is a split of the image into 4 parts. Each part will contain only a part of the 1/4 image. The features link to the position of the word in the document and the sequential position is used (i.e. what is the position of the words in compared to the text extracted with OCR). The feature Bounding box of each word corresponds to the position of the words but also the positions of each part of the image. The pre-training objectives of LayoutLMV2 are three Masked Visual-Language Modeling (MVLM), Text-Image Alignment (TIA), and Text-Image Matching (TIM). 
LayoutLMv3 \cite{LayoutLMv3} uses pretty much the same features as LayoutLMv2, their difference is image features.
LayoutLMv3  does not split the image into 4 but into a sequence of uniform P × P patches (i.e P is 16 in the pre-trained model). LayoutLMv3 pre-training objective is Masked Visual-Language Modeling (MVLM), Masked Image Modeling (MIM) and Word-Patch Alignment (WPA). 

Compared to  the LayoutLM-based method  Donut\cite{Donut} is only the image feature based. is not using the information extracted from the OCR like the majority of models but it is based only on the image. The authors demonstrated the efficiency of both correct answer detection and textual extraction compared to the classical OCR which often makes errors for handwritten words or which is not well visible. 
To achieve these results they first learned "how to read" in the model, and they used millions of documents for this task. They parse the document from top-left to bottom-right, their goal was to minimize the cross-entropy of the next token prediction based on the previous context. After learning how to read, they learned how to understand it. For this, they used tasks like classification, document information extraction, and document visual question answering.

To compare the capabilities of pre-trained methods with no-pre-trained models, we have taken QANet \cite{qanet} and Qalayout \cite{qalayout}. 

QANet\cite{qanet} is a question-answer model based only on text. It has shown that the unique use of attention mechanisms with convolutional models allows it to be fast and to have better performance compared to recurrent neural networks (RNNs).

QALayout is an improvement of QANet. It uses in addition to the features of QANet, other features used like images and layout information. In addition to this features uses a co-attention mechanism to better use the attention mechanism on several modalities.

\subsection{Evaluation Metrics}

\begin{table*}[tbh]
\renewcommand{\arraystretch}{1.0}
\centering
\begin{tabular}{|c|c|c|c|c|c|c|c|c|c|c|}
\hline 
  \multicolumn{2}{|c|}{Method} & Pre-trained & Param &   \multicolumn{2}{c|}{Time (second)} & \multicolumn{3}{c|}{CHIC}  & \multicolumn{1}{c|}{DocVQA} \\
\hline
\multicolumn{2}{|c|}{}& & & GPU & CPU  &  F1 & EM  & ANLS    &   ANLS  \\
\hline
\multicolumn{2}{|c|}{Human}& &  & \multicolumn{2}{c|}{24}&  83.78 & 79.14 & 90.25  & 98.1  \\
\hline
\multicolumn{2}{|c|}{Donut}& Yes & 176 M  & 1.23& 22.60  &  48.19 & 44.8 & 53.36 &  71.66\\
\hline
\multicolumn{2}{|c|}{LayoutLMv3}& Yes & 133M & 0.39 $+ \alpha$ & 1.94 $+ \alpha$ &  63.16 & 59.85 & 68.01   & 78.35 \\
\hline
\multicolumn{2}{|c|}{LayoutLMv2}& Yes & 200M & $0.21  + \alpha$&  1.65  $+ \alpha$ &  65.67 & 61.50 & 69.69   &  73.41 \\
\hline
\multicolumn{2}{|c|}{QALayout}& No & 10M & $0.25 + \alpha$ & 1.80 $+ \alpha$ & 59.05 & 46.27 & 63.19  &   53.60 \\
\hline
\multicolumn{2}{|c|}{QANet}& No & 2M  & $0.04 + \alpha$ &  0.16 $+ \alpha$ &    36.98 & 26.08  & 39.12   & 48.90  \\

\hline
\end{tabular}
\caption{This table describes the baseline methods, a number of parameters, and average processing time per question. The \textbf{$\alpha$} is average time for the OCR to extract the text from the image. In our context we used Abby OCR which takes on  2 seconds.  }\label{tab:SOTACOQ}
\end{table*}

To evaluate the performance of the models we select some metrics Exact match, Anls, F1 and IOU.
We can separate these metrics into two categories, strict metrics don't allow any errors in the answer (e.g. the Exact Match score), and soft metrics take errors such as missing a word in the answer, etc., into account. We have chosen these metrics because each of them addresses a specific need, and together they give an overall picture of the performance of the methods.

Average normalized Levenshtein  (anls) measures the distance between two  characters $(q_{k},p_{k})$ see the equation \ref{eq:anls}.
$NL(q_{k},p_{k})$ is the Normalized Levenshtein distance between ground truth (GT) and the prediction .  Then a threshold $\tau = 0.5$ to filter NL values larger than $\tau$ by returning a score of 0.
\begin{equation}\label{eq:anls}
\begin{split}
ANLS =  \max_{1..3} s(q_{k},p_{k}) \\
s(q_{k},p_{k}) = \Bigg\{
\begin{split}
(1-NL(q_{k},p_{k}) \quad \textbf{if} \quad NL(q_{k},p_{k})  < \tau \\
0                    \quad \quad    \textbf{if} \quad NL(q_{k},p_{k})  >= \tau
\end{split}
\end{split}
 \end{equation}

\begin{equation}\label{eq:f1score}
\begin{split}
precision = \frac{1 * same\_word}{tail(p_{k})}  \quad 
recall = \frac{1 * same\_word}{tail(q_{k})}  \\
f1 = \frac{2 * precision * recall}{(precision + recall)} 
\end{split}
 \end{equation}
 
The metric F1-SCORE is described in \ref{eq:f1score}. Where same\_word counts the number of similar words between GT and the prediction.\\
The exact match is a score of either 0 or 100, if the predicted answer matches exactly the expected answer then it gives 100 otherwise 0.

\subsection{CHIC experiments}
In the table, we mention the baselines tested on CHIC but also on Docvqa. We find first the name of the method used, the number of parameters of this method, and then their execution time on GPU and CPU in seconds (i.e. we have described the type of machine used for these experiments). Finally, we have the metrics of these methods on the CHIC dataset and on DocVQA.

We start our analysis with time performance, and finding the right compromise between execution time and performance is important in  an industrial context.
To get this metric we took the average response time provided by each method on both the GPU and CPU of the same machine. 
We chose GPU and CPU because in inference it is possible that some companies will operate these methods on CPU.
The \textbf{$\alpha$} means the execution time of an OCR to provide the extracted text in the image. To extract information the method needs text, they first extract the text with OCR and then uses this text and other features to predict an answer. 
The \textbf{$\alpha$} parameter varies and depends on the type of OCR used, in our context, we used Abby FineReader and on average the time taken to extract the text from an image is 2 seconds.  
The performance in time obtained shows that the fastest method is QANet and the longest is Donut. On the CHIC dataset, the method with the best promise between time performance and the most correct answers is LayoutLMv2.

When we analyze the recurrent errors on CHIC with these methods, we notice that they can be grouped into three main errors. the first concerns errors due to the OCR. These errors can be divided into two, the first is an error due to false extraction because a word is not well read by OCR. Other errors are due to the order of the words extracted by the OCR, this order has a substantial impact on the prediction. For example, if in the document we have a word Date: 12/02/2003, it is possible that sometimes the OCR reads the Date then a word in step etc and the final order is Date F72535 01/01/2004 358 12/02/2003. Finally, it will choose a word that is next to the word date so sometimes it takes the wrong date or just other numbers. 
The other type of error is confused with another value because they have the same keyword search next to it, for example, if the question is what is the job number? in some cases because the predicted answer is the first word that the keyword job, it may be job ref AF12321 whereas job no may be a little further down the text. 

\subsection{ Experimental setup}
All our experiments were carried out on the same machine. This machine contains 4 nvidia rtx A6000 GPUs 48GB and has 256 GB ram and 4 TV of SSD disk. Each experiment was performed on a single GPU. The methods we tested had different parameter sizes and different memory usage capacities. For this aspect we chose a batch size of 32 for the QANet and QALayout methods, then a batch size of 16 for the LayoutLMv3 and LayoutLMv2 methods and finally a batch size  of 1 for Donut. To train these methods we used the same Adam optimizer with a learning rate of 1e-3 at the beginning which decreases as the performance stagnates. We chose a number of epochs 200 and added an early stop  5.

\section{Conclusion}

We propose a new public dataset for visual question answering on corporate documents. We have taken all the requirements expected from a corporate dataset to propose this dataset.  CHIC contains 774 documents in several languages but also with different layouts (i.e. semi-structured, unstructured and structured document). From these documents we annotated 3384 questions and answers in various languages. These questions are also diverse as we find questions requiring an understanding of numerical value links (e.g. what is the maximum amount in the document?), links between different words in a document (e.g. what is the amount in Table I  line 3?) and other questions involving key-value understanding (e.g. what is job number?).  
This dataset will allow comparing different models on the visual question answering for corporate documents.

\section*{Acknowledgments}
This research has been funded by the LabCom IDEAS under the grand number ANR-18-LCV3-0008 , by the French ANRT agency (CIFRE program), and by the YOOZ company.




\bibliographystyle{splncs04}

\end{document}